# Parameter Sensitivity Analysis in Zinc-Ion Batteries: A Study on Ionic Conductivity, Current Density, and Electrode Properties

Roya Rajabi, Shichen Sun, Booker Wu, Jamil Khan, Kevin Huang

Department of Mechanical Engineering, University of South Carolina, Columbia, South Carolina 29208, United States of America

## Abstract

This study presents a comprehensive Multiphysics model for zinc-ion batteries (ZIBs), incorporating electrochemical aspects. The model integrates the mass transport of $Zn^{2+}$ ions, charge transfer, and solid diffusion to predict performance parameters like cell potential, and energy density. Significant research has focused on enhancing battery performance by optimizing components of battery to improve parameters such as ionic conductivity and exchange current density and capacity. In this study, we present a model-based investigation of zinc-ion batteries, examining the impact of these parameters. Our findings reveal that at low current densities, raising of ionic conductivity beyond 1.3 S/m and exchange current density above 0.13 mA/cm² do not yield substantial improvements in capacity. These insights underscore the importance of identifying performance thresholds in the development of next-generation batteries.

## Introduction

Rechargeable aqueous zinc-ion batteries (ZIBs) have emerged as promising candidates for large-scale energy storage applications, primarily due to their inherent safety features and lower costs compared to traditional lithium-ion batteries [1, 2]. Despite these advantages, the limited energy density of ZIBs remains a significant barrier to their commercialization. Energy density, a critical metric for battery performance, is determined by the product of specific volumetric capacity and working voltage. Recent research efforts have increasingly focused on optimizing these factors to develop aqueous ZIBs with higher working voltages and capacities, aiming to bridge the gap towards their practical implementation.

The cell voltage in aqueous zinc-ion batteries (ZIBs) is determined by the average reaction potential difference between the cathode and anode, which is influenced by the concentration of zinc ions [3]. Consequently, the development of cathode materials capable of operating in a high-voltage region is essential, given that zinc metal is predominantly used as the anode material. Recent research has focused on enhancing the working voltage of aqueous ZIBs through various strategies. These include electrolyte modification techniques to enable high-voltage operation, as well as modifications to the zinc anode to improve the practical applicability of high-energy-density ZIBs. Additionally, advances in cathode active materials are critical for achieving batteries with increased energy density [4-6]. Understanding the impact of each individual parameter on battery performance is crucial for effective modifications. However, isolating the effects of specific parameters poses a significant challenge. Numerical modeling and simulation have emerged as essential tools to complement experimental studies, offering detailed insights into parameters that are otherwise difficult to measure directly and enabling controlled analysis of specific variables [7].

Modeling of batteries started with developing numerical model of the lithium ion battery with two electrodes and separator by Fuller et al [8], which simplified later by Doyle and Newman [9]. Later on, they presented the model for specific capacity against discharge rate [10]. The model for the decay of the capacity with cycle number and side reactions and potential loss as function of the film thickness in accordance with Faraday's law were developed [11]. Accuracy of the models depend on the cycling conditions can be well verified with experimental data [12]. Next, general energy balance extended for the energy balance of the insertion battery systems, and then the same model were used to predict the effect of side reactions and the thermal behavior of the system [13, 14].

This paper presents a mathematical formulation and numerical modeling of zinc-ion batteries using COMSOL Multiphysics, building upon existing models originally developed for lithium-ion batteries. The novelty lies in adapting and applying this framework specifically to zinc-ion systems to replicate their behavior and validate the model against experimental data. The study focuses on performing a sensitivity analysis to assess the impact of key parameters such as ionic conductivity, electrode thickness, and exchange current density on battery capacity. The findings reveal that optimizing battery performance does not necessarily require materials with the highest ionic conductivity or exchange current density. At low current densities, these properties have a limited effect, underscoring the critical role of operating conditions rather than solely relying on material enhancements. This conclusion emphasizes that tailoring running conditions is crucial for improving the practical performance of zinc-ion batteries.

**Experimental Section**

The experiments in this work were performed on a coin cell zinc-ion battery with a configuration consisting of a zinc foil anode and a solid hydrogel electrolyte (CSAM-T) made from carboxymethyl chitosan (CMCS), acrylamide monomer (AM), and a photo initiator soaked in zinc triflate for 24 hours [6]. The cathode was a pre-inserted zinc into vanadium oxide ($Zn_{0.1}V_2O_5 \cdot nH_2O$) [15]. The performance of the battery with configuration of (Zn/hydrogel/ZnVO) was assessed in CR2032 coin cells. The battery underwent galvanostatic charging and discharging cycles using a LAND battery testing system (CT2001A) at both low and high current densities was collected to validate the model.

**Zinc-ion electrochemical model**

In a battery, critical processes such as diffusion, migration, and charge transfer are captured using governing equations like Fick's Law, the Nernst-Planck equation, and the Butler-Volmer equation within a 1-D framework. Key parameters, including ionic conductivity, electronic conductivity, and diffusion coefficients, are derived from a combination of experimental data and established literature sources. The model's accuracy is validated by comparing simulated voltage-capacity curves with experimental results. By precisely fitting zinc concentration profiles at electrode interfaces, the model demonstrates strong agreement with experimental data, providing deep insights into the battery's performance under different operating conditions. This study advances the understanding of zinc-ion battery dynamics and supports the optimization of their design and operation, contributing to the development of more efficient and sustainable energy storage solutions.

A typical battery consists of two current collectors, a positive electrode, a separator, and a negative electrode. The porous electrodes are infiltrated with a zinc salt solution, which acts as the electrolyte. Figure 1a presents a schematic representation of a coin cell zinc-ion battery along with its components. The model's 2-D and 1-D views are depicted in Figures 1b and 1c, respectively.

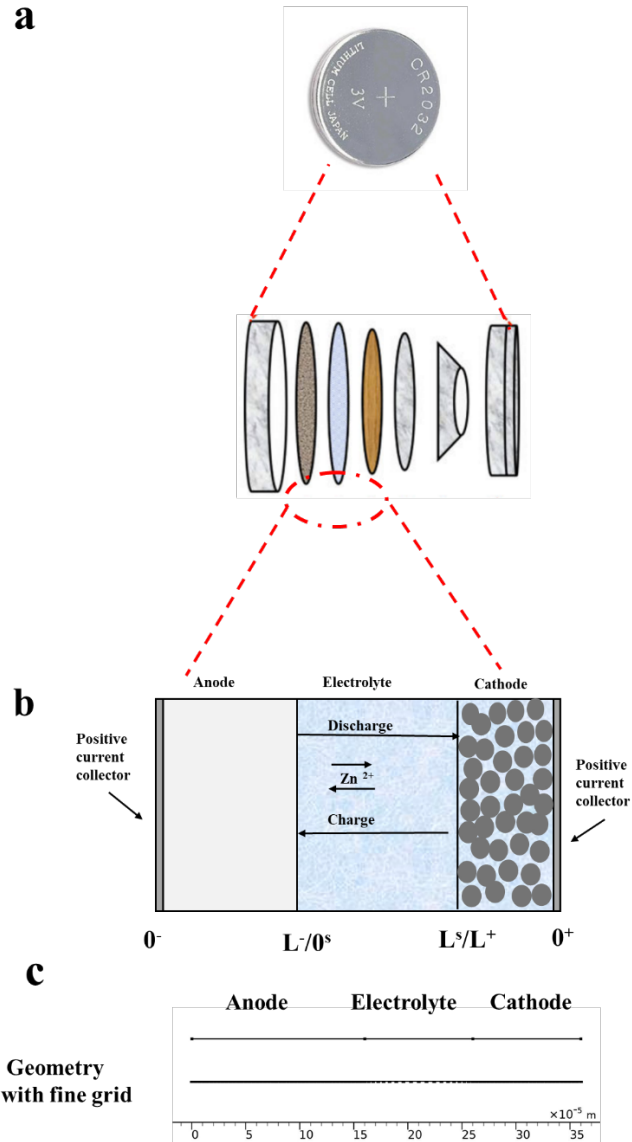

Figure 1: a) Schematic of a zinc ion battery b) Schematic of the 1-D model

In this study, the positive electrode is modeled as a porous structure to account for electrolyte diffusion within its matrix. The negative electrode is of Zn metal, where electrolyte diffusion is negligible. Therefore, in the numerical model, the negative electrode is treated as a non-porous entity, as electrochemical reactions predominantly occur at the electrode-electrolyte interface,

A 1-D model is employed to simulate battery behavior, with a fine mesh (95 elements) which edge effects in the length and height dimensions are neglected. The electrochemical processes in the battery involve electronic current conduction, initiated by applying current through an external circuit. These current drives charge transport in both the electrodes and the electrolyte. The resulting material transport in the electrolyte generates a concentration gradient, highlighting the significance of ionic conductivity and concentration over potential.

Assuming charge neutrality within solid spherical particles of radius $R_p$, the material balance for Zn-ions in active material particles is governed by Fick's second law in the radial direction:

$$\frac{\partial C_s(r,t)}{\partial t} = D_{s,i} \frac{1}{r^2} \frac{\partial}{\partial r}\left(r^2 \frac{\partial C_s(r,t)}{\partial r}\right) \tag{1}$$

where $C_s(r,t)$ donates the zinc-ion concentration in electrodes at radial coordinate r, $D_s$ is the solid diffusion coefficient of Zn-ions. The boundary conditions of Eq. (8.1) are given by:

$$\left.\frac{\partial C_s(r,t)}{\partial r}\right|_{r=0} = 0 \tag{2}$$

$$\left.\frac{\partial C_s(r,t)}{\partial r}\right|_{r=Rp} = \frac{j(t)}{D_s} \tag{3}$$

with j(t) as the molar flux of Zn-ions on the particle surface.

Considering the changes of Zn-ions in electrolyte due to the concentration difference-induced diffusion and electrolyte current, the electrolyte dynamics can be defined by:

$$\frac{\partial C_e}{\partial t} = \frac{\partial}{\partial x}\left(D_{e,eff} \frac{\partial C_e}{\partial x}\right) + \frac{(1-t_0^+)}{\varepsilon_e F}\left(\frac{\partial i_e}{\partial x}\right) \tag{4}$$

where $C_e$ and $i_e$ denote the Zn-ion concentration and the current in the electrolyte, $D_{e,eff}$ is the effective electrolyte diffusion coefficient, $t_0^+$ is the transference number for the cations, $\varepsilon_e$ is the volume fraction of the electrolyte and F is the Faraday's constant. The spatial change of $i_e$ and effective diffusion coefficient can be described by:

$$\frac{\partial i_e}{\partial x} = a_i F J_i \tag{5}$$

$$D_{eff,i} = D_{e,i} \varepsilon_i^{bruggi} \tag{6}$$

where a is the specific interfacial area, and $brugg$ is the Bruggeman coefficient and is 1.5 for ZIB (Song, Zhang et al. 2024).

The boundary conditions at current collectors and electrodes and hydrogel electrolyte are as follows,

$$\frac{\partial C_e}{\partial x}\bigg|_{x=0^-} = \frac{\partial C_e}{\partial x}\bigg|_{x=0^+} = 0 \tag{7}$$

$$\varepsilon_e^- D_{e,eff} \frac{\partial C_e}{\partial x}\bigg|_{x=L^-} = \varepsilon_e^s D_{e,eff} \frac{\partial C_e}{\partial x}\bigg|_{x=L^s} \tag{8}$$

$$\varepsilon_e^s D_{e,eff} \frac{\partial C_e}{\partial x}\bigg|_{x=L^s} = \varepsilon_e^+ D_{e,eff} \frac{\partial C_e}{\partial x}\bigg|_{x=L^+} \tag{9}$$

$$C_e(L^-, t) = C_e(L^s, t) \tag{10}$$

$$C_e(L^s, t) = C_e(L^+, t) \tag{11}$$

where $0^-$ and $0^+$ correspond to the boundaries between current collectors and electrodes, $L^-/L^s$ and $L^s/L^+$ corresponds to the boundaries between separator and electrodes.

$$\sigma_{eff,s} \frac{\partial^2 \varphi_s}{\partial x^2} = aFJ \tag{12}$$

where $\sigma_{eff}$ is the effective electrode conductivity.

The overpotential of the cell $\eta$ is given by:

$$\eta = \phi_s - \phi_e - U\left(\frac{C_s}{C_{s,max}}\right) - FR_fJ \tag{13}$$

where $C_{s,max}$ is the maximum zinc-ion concentration of the electrode and $R_f$ is the resistance between the electrode and electrolyte (SEI film resistance) which is neglected here. As shown in Eqn. (8.14), the electrode open-circuit potential depends on zinc-ion concentration on the surface of the electrode.

The relationship between the molar flux, j, and the overpotential is given by Butler-Volmer kinetics:

$$J = \frac{i_0}{F}\left[\exp\left(\frac{(1-\alpha)F}{RT}\eta\right) - \exp\left(-\frac{(1-\alpha)F}{RT}\eta\right)\right] \quad (14)$$

where $\alpha$ charge transfer coefficient and the exchange current density $i_0(x,t)$ is defined by:

$$i_0 = kC_e^\alpha (C_{s,max} - C_s)^{(1-\alpha)} C_s^\alpha \quad (15)$$

where k is the reaction rate coefficient which can be obtained by:

$$k_i = \frac{i_{0,i}}{F\sqrt{C_{s,i} * (C_{s,max,i} - C_{s,i})}} \quad (16)$$

where i can be positive or negative.

At the interface of current collector and the positive electrode, the charge flux is equal to the current density applied to the cell.

$$-\sigma_{eff} \frac{\partial \varphi_e}{\partial x}\bigg|_{x=0} = I_{app} \quad (17)$$

where $I_{app}$ is the applied current density.

Zinc-ion batteries incorporating a semisolid CSAM-T electrolyte are modeled using equations (1)-(17). To reduce computational cost, a 1-D model was developed, which is versatile enough to represent not only a specific cell configuration but also other cells sharing the same electrochemical properties. This approach ensures computational efficiency while maintaining accuracy in simulating the battery's performance across different designs within the same chemistry [3, 18, 19].

**Parameter Benchmarking for zinc-ZnVO cell**

The focused parameters in this work include geometric parameters, transport parameters, kinetics parameters, and concentration parameters which were obtained through experiment measurement, previous works, and fitting. The reaction mechanism in the anode and cathode can be expressed as below:

Anodic process ($H^+$ and $Zn^{2+}$ insertion/reaction):

$$Zn \rightarrow Zn^{2+} + 2e^-$$

Cathodic process: ($H^+$ and $Zn^{2+}$ insertion/reaction and zinc hydroxide formation/dissolution)

$$yZn^{2+} + Zn_xV_2O_5 + 2ye^- \rightarrow Zn_{x+y}V_2O_5$$

Some parameters like the dimension of the cell and ionic conductivity of the electrolyte and electronic conductivity of the electrode have been obtained from experiment as listed in table 1.

Table 1: List of quantities obtained from experiment

| Quantity | Units | Value |
|---|---|---|
| **Experimental measurement** | | |
| Anode Thickness (Ln) | μm | 200 |
| Cathode Thickness (Lp) | μm | 200 |
| Electrolyte Thickness (Ls) | μm | 100 |
| Electrode Surface Area (a) | μm² | |
| Electrolyte Ionic Conductivity ($\sigma$) | S/m | 1.3 |
| Zinc Electrical Conductivity | S/m | 1.69×10⁷ |
| Cathode Active Material Volume Reaction ($\varepsilon_p$) | - | 0.69 |
| Filler Volume in Cathode ($\varepsilon_{f-p}$) | - | 0.41 |
| **Previous Works** | | |
| Zn-diffusivity Negative Electrode (Ds$_{neg}$) | $m^2/s$ | 1.96e-9 |
| Zn-diffusivity Positive Electrode (Ds$_{pos}$) | $m^2/s$ | 7.27e-7 |
| Particle Radius Negative Electrode (Rp$_{neg}$) | μm | 8 |
| Anode Exchange Current Density (i$_{neg}$) | mA/cm² | 9.5e-2 |

**Verification of the model with experimental data**

The mathematical model employed is a 1-D isothermal representation consisting of three sequential segments corresponding to the negative electrode, the separator, and the positive electrode. This model is designed to elucidate the influence of critical parameters and to assess the performance of zinc-ion batteries under various conditions. It is derived from the framework established by J. Newman and colleagues [19]. The model's accuracy was validated through experimental measurements conducted at current densities of 1 A/m² and 25 A/m² over approximately 19,000 seconds until the cell voltage declined to 0.2 V. Ion concentration at the interfaces was determined using a fitting method, and the results demonstrated strong concordance between experimental data and the model predictions, as shown in Fig. 2. Input data based on the validated model are presented in Tables 3 and 4 for the two different applied currents.

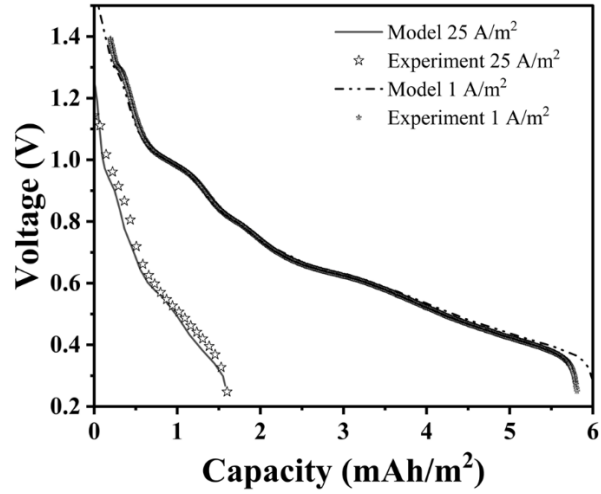

Figure 2: Capacity- voltage curve of the model compared with experimental data at 1A/m².

Table 2: List of quantities obtained from optimization of modeling and experimental after validation at 1 A/m²

| Quantity | Units | Value |
|---|---|---|
| Initial Concentration Positive Active Electrode Material (Cs0pos) | mole/m³ | 250 |
| Max Solid Phase Concentration, Positive Electrode (Cspos-max) | mole/m³ | 9800 |
| Exchange Current Density, Positive Electrode ($i_{pos}$) | A/m² | 0.2 |

Table 3: Table 3: List of quantities obtained from optimization of modeling and experimental after validation at 25 A/m²

| Quantity | Units | Value |
|---|---|---|
| Initial Concentration Positive Active Electrode Material (Cs0pos) | mole/m³ | 250 |
| Max Solid Phase Concentration, Positive Electrode (Cspos-max) | mole/m³ | 2800 |
| Exchange Current Density, Positive Electrode ($i_{pos}$) | A/m² | 1.3 |

The fitting results reveal that increasing the current density from 1 A/m² to 25 A/m² corresponds to a reduction in the maximum concentration of zinc ions, which decreases from 9,800 mol/m³ to 2,800 mol/m³. This trend indicates that higher current densities enhance species transport, thereby accelerating the reaction rate. Consequently, this leads to an increase in the exchange current density, which varies from 0.2 A/m² to 1.3 A/m².

**Parameter sensitivity analysis**

The material properties are representative of a typical zinc-ion battery configuration. The electrolyte comprises a hydrogel with a 1 M zinc triflate solution, while the anode is constructed from zinc foil and the cathode material is $Zn_{0.1}V_2O_5$. The electrolyte conductivity was experimentally determined to be 13 mS/cm. The equilibrium potential of the positive electrode was obtained through galvanostatic intermittent titration technique (GITT) and validated with experimental measurements.

The validated model was employed to analyze the impact of C-rate, electrode thickness, and exchange current density on the voltage-capacity profile of the zinc-ion battery. An increase in the C-rate results in a reduced capacity and a shorter voltage-capacity profile, leading to a lower energy density, as illustrated in Figures 3a and 3b. For a current density of 12.5 A/m², the energy density is 246.98 Wh/m². As the current density increases (12.5, 25, 50, 75, and 100 A/m²), the state of charge diminishes, thereby reducing the energy density (246.98, 236.29, 51.6, 15.7, and 0.76 Wh/m², respectively), as shown in Figure 3b. A linear relationship is observed between increasing current density and decreasing energy density; specifically, doubling the current density results in a 4% capacity loss, while increases of 200%, 300%, and 700% lead to capacity reductions of 79%, 93%, and 99.99%, respectively.

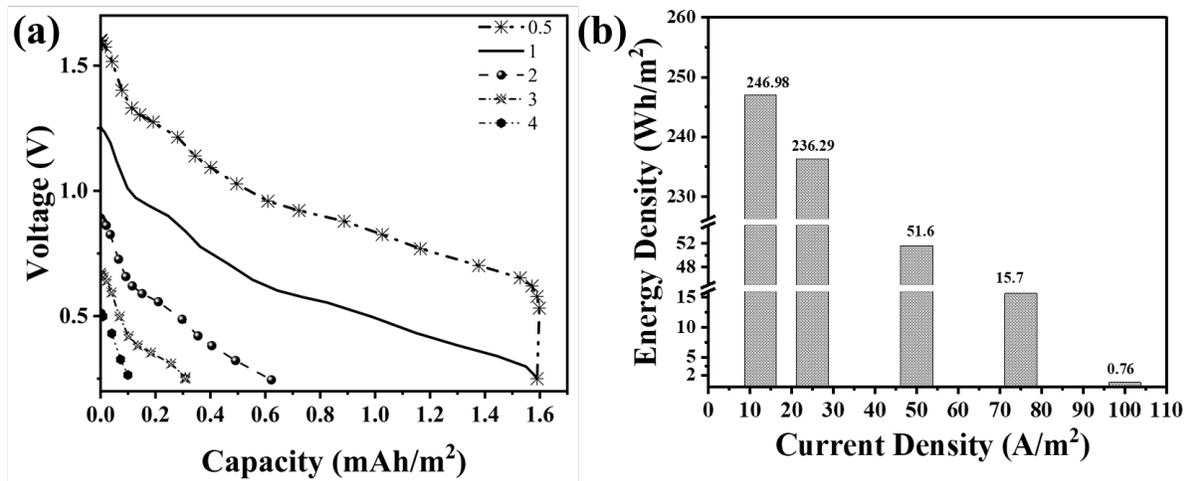

Figure 3: Voltage- capacity profile of the zinc ion battery with CSAM-T hydrogel electrolyte with 1.3 S/m ionic conductivity at 0.5C, 1C, 2C, 3C, 4C with 25 A/m² current density b) Energy density of the battery at 0.5C, 1C, 2C, 3C, 4C with 25 A/m² current density

Improving battery capacity by enhancing the ionic conductivity, particularly in batteries using polymer electrolytes, remains a significant challenge [20]. Understanding when and to what extent increased ionic conductivity positively impacts battery capacity is crucial. Figure 4a illustrates that at a current density of 25 A/m², an ionic conductivity exceeding 1.3 mS/cm in a zinc-ion battery with ZnVO as the cathode and a hydrogel electrolyte does not significantly affect the battery's capacity and as a result energy density. However, the impact of ionic conductivity varies with different C-rates. As depicted in Figure 4b, at higher current densities above 25 A/m², the effect of ionic conductivity becomes more pronounced. For instance, increasing the current density from 25 to approximately 60 A/m² results in a more significant capacity difference. This indicates that higher C-rates necessitate faster ion movement, making high ionic conductivity crucial for devices requiring rapid charge or discharge. According to Equation (15), elevated ionic conductivity leads to increased molar flux and concentration gradient. It is important to note that the threshold at which further increases in ionic conductivity do not produce additional benefits varies depending on the battery configuration and current density.

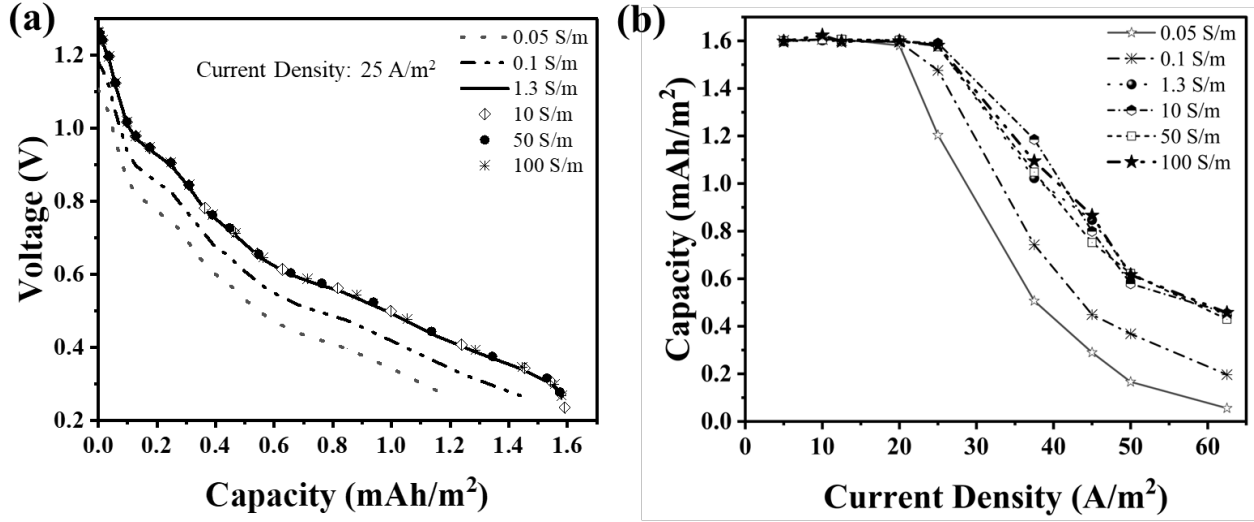

Figure 4: a) Effect of ionic conductivity on the voltage capacity profile b) Effect of ionic conductivity at different current density on capacity

Increasing the thickness leads to higher active material in the electrode of the battery and theoretically causes a higher capacity. However, it has been found that thicker electrodes lead to uneven heating, which degrades the active material and reduces capacity. Also ions have longer diffusion paths, and limiting access to active material [21]. Altering the thickness of the anode and cathode can impact battery performance, particularly if it results in increased resistance. In the mathematical model, increasing the thickness of the cathode, which corresponds to a greater amount of active material as indicated by Equation (18), leads to higher initial capacity, as demonstrated in Figure 5a. However, this improvement in initial capacity may be offset by increased internal resistance, potentially reducing overall performance. Thus, while a thicker cathode enhances capacity initially, careful consideration of resistance is necessary to optimize battery performance.

$$Initial\ Capacity = Cs_{max} * SoC_{max} * \varepsilon * L_p * F \tag{18}$$

However, in the case of the anode electrode, where zinc foil with effectively infinite conductivity is used, increasing the anode thickness does not significantly affect battery performance, as illustrated in Figure 5b.

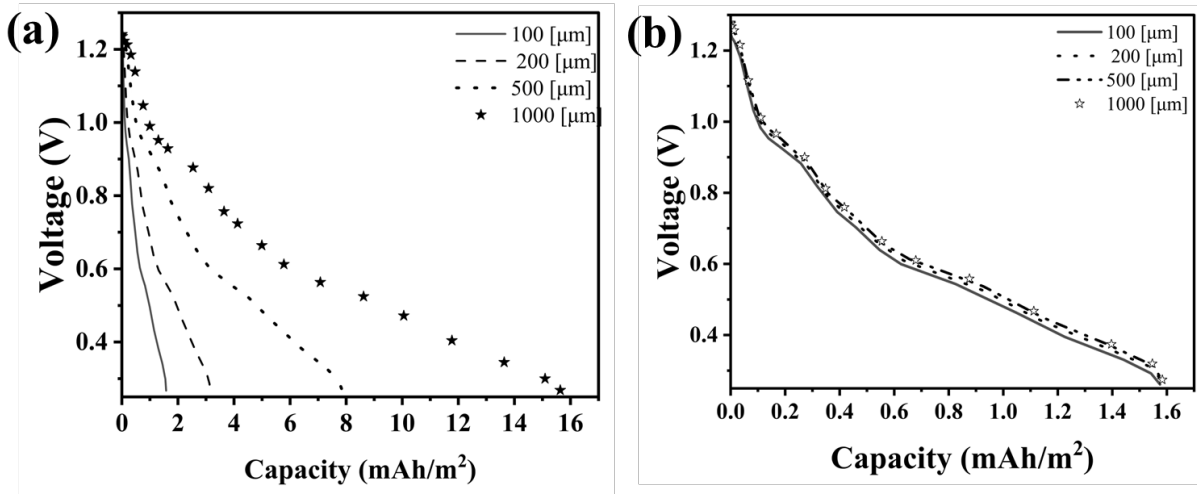

Figure 5: Effect of the electrode's thickness on the battery's voltage-capacity profile with 1.3 S/m conductivity and 25 A/m² applied current a) cathood's thickness b) anode's thickness

However, this effect may vary with changes in current density. At lower current densities, increasing the thickness of the active material significantly enhances capacity. Conversely, at higher current densities, the impact of thickness diminishes due to the rapid ion movement. The influence of the active material on battery performance is also evident when modifying the proportion of inactive filler in the cathode. Increasing the filler content results in a reduction in initial capacity, underscoring the critical role of active materials in the cathode. This relationship is illustrated in Figure 6b.

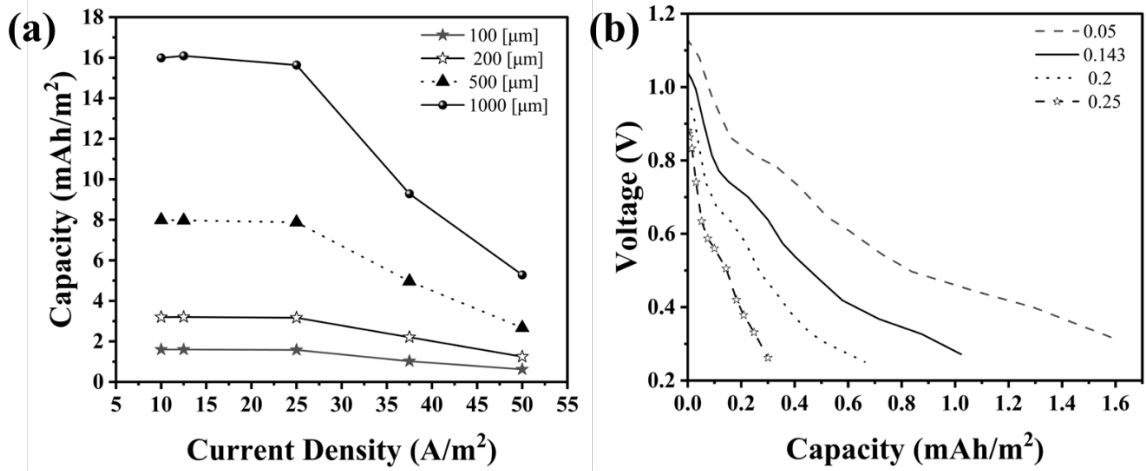

Figure 6: a) effect of the cathode 's thickness at different current densities with electrolyte's ionic conductivity of 1.3 S/m, b) Effect of filler percentage in cathode in a coil cell battery with electrolyte's ionic conductivity of 1.3 S/m and at 25 A/m²

A higher exchange current density means that extensive oxidation and reduction reactions occur while a low value indicates the opposite [22]. An increased exchange current density enhances both capacity and voltage. As illustrated in Fig. 7a, for a battery with an electrolyte ionic conductivity of 1.3 S/m and a current density of 25 A/m², varying the exchange current density by ±10 times results in approximately ±45% variation in initial voltage. However, a lower exchange current density leads to a significant reduction of 84% in initial capacity. At high current densities (1.3 mA/cm²), changes in exchange current density have minimal impact on capacity. In contrast, at an exchange current density of 0.013 mA/cm², increasing the current density drastically reduces capacity. For an exchange current density of 0.13 mA/cm², capacity remains stable up to 25 A/m², after which further increases in current density led to a marked decline in capacity.

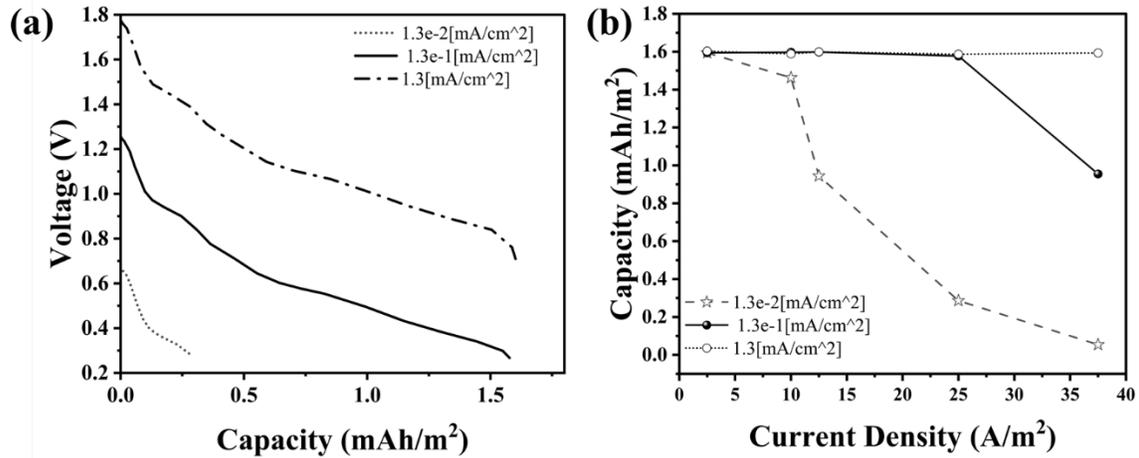

Figure 7: a) Effect of exchange current density on the voltage-capacity performance of a battery, b) Effect of exchange current density at different current density with electrolyte ionic conductivity of 1.3 S/m

**Conclusion**

This study provides a comprehensive analysis of the performance characteristics and key parameters affecting zinc-ion batteries (ZIBs) through both experimental and mathematical modeling approaches. Our results demonstrate that variations in current density, electrode thickness, and exchange current density significantly influence the voltage-capacity profile and overall energy density of ZIBs. The 1-D model employed effectively captures the essential electrochemical processes, including charge transport and material diffusion, providing valuable insights into the battery's behavior under various operational conditions. The experimental validation of the model reveals a strong correlation between simulated and experimental data, confirming the model's reliability for predicting battery performance.

Key findings include:

*Current Density Effects:* Increasing the current density from 1 A/m² to 25 A/m² results in a substantial decrease in the maximum concentration of zinc ions, leading to reduced energy density and shorter voltage-capacity profiles.

*Electrolyte Ionic Conductivity:* While higher ionic conductivity can enhance battery performance, its impact becomes less pronounced at current densities above 25 A/m². This suggests that optimizing ionic conductivity is crucial for applications requiring rapid charge or discharge, but its benefits are limited under high current conditions.

*Electrode Thickness:* Increasing the thickness of the cathode enhances the initial capacity due to the greater active material volume, although this effect is constrained by the resistance and efficiency of the charge transfer. Conversely, the thickness of the zinc foil as anode, with its inherent high conductivity, does not significantly affect performance.

*Exchange Current Density:* A higher exchange current density correlates with increased capacity and voltage. Significant changes in exchange current density led to notable variations in initial voltage and capacity. At lower exchange current densities, increased current density adversely affects battery capacity, while at higher densities, the impact on capacity stabilizes until a threshold is exceeded.

Overall, this study advances the understanding of ZIB dynamics and provides practical guidelines for optimizing battery design and operation. The findings underscore the importance of balancing key parameters to achieve desired performance outcomes and support the development of more efficient and sustainable energy storage technologies.